\documentclass[%
twocolumn,
letterpaper,
superscriptaddress,
showpacs,
preprintnumbers,
amsmath,amssymb,
aps,
prc
floatfix,
longbibliography
]{revtex4-1}

\usepackage{graphicx}
\usepackage{dcolumn}
\usepackage{bm}
\usepackage{amsfonts}
\usepackage{amsmath}
\usepackage{cases}
\usepackage{txfonts}
\usepackage{amssymb}
\usepackage{mathrsfs}
\usepackage{graphicx}
\usepackage{multirow}
\usepackage{float}

\newcommand{\snn}{\sqrt{s_\text{NN}}}
\newcommand{\pt}{p_T}
\newcommand{\gc}{GeV/$c$}
\newcommand{\bv}[1]{\textbf{\emph{{}#1}}}
\newcommand{\ffps}[1]{E_{{}#1}\frac{d^3\bv{N}_{{}#1}}{d\bv{p}^3_{{}#1}}}

\newcommand{\npart}{\langle N_{\textrm{part}}\rangle}
\begin{document}

\title{Beam energy dependence of (anti-)deuteron production in Au+Au collisions at RHIC}
\affiliation{Abilene Christian University, Abilene, Texas   79699}
\affiliation{AGH University of Science and Technology, FPACS, Cracow 30-059, Poland}
\affiliation{Alikhanov Institute for Theoretical and Experimental Physics, Moscow 117218, Russia}
\affiliation{Argonne National Laboratory, Argonne, Illinois 60439}
\affiliation{Brookhaven National Laboratory, Upton, New York 11973}
\affiliation{University of California, Berkeley, California 94720}
\affiliation{University of California, Davis, California 95616}
\affiliation{University of California, Los Angeles, California 90095}
\affiliation{University of California, Riverside, California 92521}
\affiliation{Central China Normal University, Wuhan, Hubei 430079 }
\affiliation{University of Illinois at Chicago, Chicago, Illinois 60607}
\affiliation{Creighton University, Omaha, Nebraska 68178}
\affiliation{Czech Technical University in Prague, FNSPE, Prague 115 19, Czech Republic}
\affiliation{Technische Universit\"at Darmstadt, Darmstadt 64289, Germany}
\affiliation{E\"otv\"os Lor\'and University, Budapest, Hungary H-1117}
\affiliation{Frankfurt Institute for Advanced Studies FIAS, Frankfurt 60438, Germany}
\affiliation{Fudan University, Shanghai, 200433 }
\affiliation{University of Heidelberg, Heidelberg 69120, Germany }
\affiliation{University of Houston, Houston, Texas 77204}
\affiliation{Huzhou University, Huzhou, Zhejiang  313000}
\affiliation{Indiana University, Bloomington, Indiana 47408}
\affiliation{Institute of Physics, Bhubaneswar 751005, India}
\affiliation{University of Jammu, Jammu 180001, India}
\affiliation{Joint Institute for Nuclear Research, Dubna 141 980, Russia}
\affiliation{Kent State University, Kent, Ohio 44242}
\affiliation{University of Kentucky, Lexington, Kentucky 40506-0055}
\affiliation{Lawrence Berkeley National Laboratory, Berkeley, California 94720}
\affiliation{Lehigh University, Bethlehem, Pennsylvania 18015}
\affiliation{Max-Planck-Institut f\"ur Physik, Munich 80805, Germany}
\affiliation{Michigan State University, East Lansing, Michigan 48824}
\affiliation{National Research Nuclear University MEPhI, Moscow 115409, Russia}
\affiliation{National Institute of Science Education and Research, HBNI, Jatni 752050, India}
\affiliation{National Cheng Kung University, Tainan 70101 }
\affiliation{Nuclear Physics Institute of the CAS, Rez 250 68, Czech Republic}
\affiliation{Ohio State University, Columbus, Ohio 43210}
\affiliation{Institute of Nuclear Physics PAN, Cracow 31-342, Poland}
\affiliation{Panjab University, Chandigarh 160014, India}
\affiliation{Pennsylvania State University, University Park, Pennsylvania 16802}
\affiliation{NRC "Kurchatov Institute", Institute of High Energy Physics, Protvino 142281, Russia}
\affiliation{Purdue University, West Lafayette, Indiana 47907}
\affiliation{Pusan National University, Pusan 46241, Korea}
\affiliation{Rice University, Houston, Texas 77251}
\affiliation{Rutgers University, Piscataway, New Jersey 08854}
\affiliation{Universidade de S\~ao Paulo, S\~ao Paulo, Brazil 05314-970}
\affiliation{University of Science and Technology of China, Hefei, Anhui 230026}
\affiliation{Shandong University, Qingdao, Shandong 266237}
\affiliation{Shanghai Institute of Applied Physics, Chinese Academy of Sciences, Shanghai 201800}
\affiliation{Southern Connecticut State University, New Haven, Connecticut 06515}
\affiliation{State University of New York, Stony Brook, New York 11794}
\affiliation{Temple University, Philadelphia, Pennsylvania 19122}
\affiliation{Texas A\&M University, College Station, Texas 77843}
\affiliation{University of Texas, Austin, Texas 78712}
\affiliation{Tsinghua University, Beijing 100084}
\affiliation{University of Tsukuba, Tsukuba, Ibaraki 305-8571, Japan}
\affiliation{United States Naval Academy, Annapolis, Maryland 21402}
\affiliation{Valparaiso University, Valparaiso, Indiana 46383}
\affiliation{Variable Energy Cyclotron Centre, Kolkata 700064, India}
\affiliation{Warsaw University of Technology, Warsaw 00-661, Poland}
\affiliation{Wayne State University, Detroit, Michigan 48201}
\affiliation{Yale University, New Haven, Connecticut 06520}

\author{J.~Adam}\affiliation{Creighton University, Omaha, Nebraska 68178}
\author{L.~Adamczyk}\affiliation{AGH University of Science and Technology, FPACS, Cracow 30-059, Poland}
\author{J.~R.~Adams}\affiliation{Ohio State University, Columbus, Ohio 43210}
\author{J.~K.~Adkins}\affiliation{University of Kentucky, Lexington, Kentucky 40506-0055}
\author{G.~Agakishiev}\affiliation{Joint Institute for Nuclear Research, Dubna 141 980, Russia}
\author{M.~M.~Aggarwal}\affiliation{Panjab University, Chandigarh 160014, India}
\author{Z.~Ahammed}\affiliation{Variable Energy Cyclotron Centre, Kolkata 700064, India}
\author{I.~Alekseev}\affiliation{Alikhanov Institute for Theoretical and Experimental Physics, Moscow 117218, Russia}\affiliation{National Research Nuclear University MEPhI, Moscow 115409, Russia}
\author{D.~M.~Anderson}\affiliation{Texas A\&M University, College Station, Texas 77843}
\author{R.~Aoyama}\affiliation{University of Tsukuba, Tsukuba, Ibaraki 305-8571, Japan}
% ...

\collaboration{STAR Collaboration}\noaffiliation

\date{\today}

\begin{abstract}
We report the energy dependence of mid-rapidity (anti-)deuteron production in Au+Au collisions at $\snn =\ $7.7, 11.5, 14.5, 19.6, 27, 39, 62.4, and 200 GeV, measured by the STAR experiment at RHIC. The yield of deuterons is found to be well described by the thermal model. The collision energy, centrality, and transverse momentum dependence of the coalescence parameter $B_2$ are discussed. We find that the values of $B_2$ for anti-deuterons are systematically lower than those for deuterons, indicating that the correlation volume of anti-baryons is larger than that of baryons at $\snn$ from 19.6 to 39 GeV. In addition, values of $B_2$ are found to vary with collision energy and show a broad minimum around $\snn =\ $20 to 40 GeV, which might imply a change of the equation of state of the medium in these collisions.
\end{abstract}

\pacs{25.75.-q, 25.75.Ld}
\maketitle


\section{\label{sec:level1}Introduction}
The main goal of ultra-relativistic heavy-ion collisions is the creation of a new state of matter, the quark-gluon plasma (QGP) in laboratories. After strongly coupled QGP was observed at the Relativistic Heavy Ion Collider (RHIC)~\cite{ADAMS2005102,ADCOX2005184,BACK200528,ARSENE20051}, attempts are being made to vary the colliding beam energy and to study the phase structure of QCD matter expressed in terms of a $T-\mu_B$ phase diagram, which is the core physics program for the Beam Energy Scan (BES) at RHIC.~\cite{Stephanov,Mohanty,Aggarwal:2010cw,PhysRevLett.105.022302,PhysRevLett.112.032302,PhysRevLett.113.092301,2018551}. 

In relativistic heavy-ion collisions, the underlying mechanism for light (anti-)nuclei production is not well understood~\cite{PhysRevC.94.034908,ALICE,CHEN20181}. One possible approach is through coalescence of (anti-)nucleons~\cite{PhysRevLett.37.667,PhysRevC.59.1585,PhysRevC.52.2004,SATO1981153,SUN2017103}. Since the binding energies of light nuclei are small ($\sim$2.2 MeV for (anti-)deuteron and $\sim$7.7 MeV for $^3$He), these light nuclei cannot survive when the temperature is much higher than the binding energy. The typical kinetic freeze-out temperature for light hadrons is around 100 MeV~\cite{PhysRevC.94.034908}, hence they might break apart and be formed again by final-state coalescence after nucleons are de-coupled from the hot and dense system. Therefore the production of these light nuclei can be used to extract important information of nucleon distributions at freeze-out~\cite{PhysRevLett.37.667,SATO1981153,PhysRev.129.836}. In the coalescence picture, the invariant yield of light nuclei is proportional to the invariant yield of nucleons :
\begin{eqnarray}\label{b2eq}
\ffps{A}&=&{B_A}{\left(\ffps{p}\right)^Z}{\left(\ffps{n}\right)^{A - Z}}\nonumber
\\&\approx&{B_A}{\left(\ffps{p}\right)^A},
\end{eqnarray}
where $A$ and $Z$ are the mass and charge number of the light nucleus under study. $\bv{p}_p$, $\bv{p}_n$, and $\bv{p}_A$ are momenta of proton, neutron, and nucleus respectively, with $\bv{p}_A = A\bv{p}_p$, assuming $\bv{p}_p\sim\bv{p}_n$. The coalescence parameter $B_A$ reflects the probability of nucleon coalescence, and is related to the local nucleon density. This coalescence approach works quite well in nuclear interactions at low energy~\cite{200022,PhysRevLett.87.262301,PhysRevC.69.024902}. The effective volume of the nuclear matter at the time of condensation of nucleons into light nuclei, also called ``nucleon correlation volume $V_{\text{eff}}$'', is related to the coalescence parameter $B_A$~\cite{CSERNAI1986223},
\begin{equation}\label{b2vsv}
{B_A}\propto V_{\text{eff}}^{1-A}.
\end{equation}
The production of light nuclei can also be described by thermodynamic models~\cite{Andronic2011203,PhysRevC.84.054916,0954-3899-38-12-124081,PhysRevC.94.034908}, in which chemical equilibrium among protons, neutrons and light nuclei is attained. Therefore, the production of light nuclei provides a tool to measure the freeze-out properties.

The production of light nuclei has been studied extensively at collision energies available at the AGS~\cite{E878,E802,E864-1,E864-2}, SPS~\cite{NA52}, RHIC~\cite{PhysRevLett.87.262301,PhysRevLett.94.122302,PhysRevC.69.034909,PhysRevC.83.044906}, and LHC~\cite{ALICE}. Studies of the $\snn$ dependence of some observables are of particular interest, because production mechanisms of light nuclei might be different at different collision energies. For example, at low energy, spectator fragmentation could be an important source for light nuclei, while at higher energy, coalescence of nucleons could become the dominant mechanism. In very high energy collisions, direct coalescence of quarks to a light nucleus could be possible. At energies below 20 GeV the $B_2$ parameter decreases with increasing collision energy~\cite{NA49,PhysRevLett.94.122302} implying the increase of the correlation volumes. At the RHIC top energy of $\snn = $ 200 GeV, the values of $B_2$ are the same for deuterons and anti-deuterons, similar to $B_2$ of deuterons from heavy-ion collisions at around $\snn = $10 GeV. In heavy-ion collisions, minima around $\snn\sim$ 20 GeV have been observed from several observables including the two-particle correlations of pions~\cite{PhysRevC.92.014904,Adare:2014qvs}, the directed flow of net-proton and net-Lambda~\cite{PhysRevLett.112.162301,PhysRevLett.120.062301}, and the $4^{\text{th}}$ order moments of net-protons~\cite{PhysRevLett.112.032302}. The results imply a dramatic change in the properties of the medium at these collision energies. The deuteron can be referred to as a system of proton-neutron correlations. The energy dependence of the deuteron elliptic flow parameter $v_2$ was published earlier by the STAR collaboration~\cite{PhysRevC.94.034908}. Above 20 GeV, $v_2$ of deuterons and anti-deuterons are found to be the same. Here we report the energy dependent results of (anti-)deuteron yields and the space-momentum correlation among nucleons. This allows an extraction of the nucleon local density at freeze-out for both nucleons and anti-nucleons. 

In this paper, a systematic study of mid-rapidity deuteron and anti-deuteron production is presented for Au+Au collisions at $\snn = $7.7, 11.5, 14.5, 19.6, 27, 39, 62.4, and 200 GeV, measured by STAR at RHIC. The paper is organized as follows. In Section ~\ref{sec:level2}, details of the STAR experiment and analysis procedure are discussed. The corrections for the detector effects and systematic uncertainties in the analysis are given in Section ~\ref{sec:level3}. In Section ~\ref{sec:level4}, the transverse momentum ($\pt$) distributions, $\pt$-integrated yields, mean transverse momenta, particle ratios, and coalescence parameters are shown. Finally, conclusions are given in Section ~\ref{sec:level5}.

\section{\label{sec:level2}Experiment and data analysis}
\subsection{\label{sec:level2-1}STAR Detector}
The results presented in this paper are obtained from the data taken with the STAR experiment~\cite{ACKERMANN2003624} in Au+Au collisions at $\snn = $7.7-200 GeV at RHIC. The 7.7, 11.5, 39, and 62.4 GeV data were collected in the year 2010, the 19.6, 27, and 200 GeV data were collected in 2011, and the 14.5 GeV data in 2014. The STAR detector has excellent particle identification capabilities. The main detectors used in this analysis are the Time Projection Chamber (TPC)~\cite{ANDERSON2003659} and the Time-of-Flight detector (ToF)~\cite{LLOPE2005306}. The TPC provides full azimuthal angle acceptance for tracks in the pseudorapidity region $|\eta| < 1$ and it also provides particle identification via the measurement of the specific energy loss $dE/dx$. It is used to identify (anti-)deuterons with transverse momenta below 1 \gc. The velocity information from the ToF detector is in addition used to identify (anti-)deuterons with transverse momenta above 1 \gc. By a combined analysis of TPC and ToF data, (anti-)deuterons can be identified up to $\pt = $ 4.8 \gc\ with statistical significance. The details of the design and other characteristics of the STAR detectors can be found in Ref.~\cite{ACKERMANN2003624}.

\subsection{\label{sec:level2-2}Event and track selection}
The primary vertex for each event is determined by finding the most probable point of common origin of the tracks measured by the TPC. As discussed in Ref.~\cite{PhysRevC.96.044904}, only those events which have the primary vertex position along the longitudinal ($V_z$) and transverse direction ($V_r$) in a certain range are selected in our analysis. These values are selected in order to achieve uniform detector performance and sufficient statistical significance of the measured observables. Table~\ref{tab:1} shows the range of $V_z$ and $V_r$ values and the total number of events after applying the vertex conditions.

\begin{table}[h!]
\caption{\label{tab:1}$V_z$ and $V_r$ conditions and total number of events for various energies obtained after all the event selection criteria are applied.}
\begin{ruledtabular}
\begin{tabular}{cccc}
$\snn$ (GeV)	&	$|V_z|<$ (cm)	&	$|V_r|<$ (cm)	&	No. of events (million)\\
\colrule
7.7	&	40	&	2	&	4\\
11.5	&	40	&	2	& 8.4\\
14.5	&	40	&	1\footnote{The center point of transverse radial position is located at $(V_x,V_y) =(0,-0.89$ cm) for 14.5 GeV.}	& 16.7\\
19.6	&	40	&	2	&	19.3\\
27	&	40	&	2	&	37.6\\
39	&	40	&	2	&	112\\
62.4	&	40	&	2	&	49.8\\
200	&	30	&	2	&	313\\
\end{tabular}
\end{ruledtabular}
\end{table}

Centralities in Au+Au collisions are defined by the number of primary charged-particle tracks reconstructed in the TPC over the full azimuth and pseudorapidity $|\eta| < $ 0.5. This is generally called the ``reference multiplicity" in STAR. The centrality classes are obtained as fractions of the reference multiplicity distribution. The events are divided into the following centrality classes: $0-10\%$, $10-20\%$, $20-40\%$, $40-60\%$, and $60-80\%$. The mean values of the number of participants ($\npart$) corresponding to these centrality classes are evaluated by Glauber model and are given in Table~\ref{tab:2} for various energies. More details on centrality and $\npart$ value estimates can be found in Refs.~\cite{PhysRevC.81.024911,PhysRevC.79.034909}.

\begin{table}[h!]
\caption{\label{tab:2}The average number of participating nucleons ($\npart$) for various collision centralities in Au+Au collisions at $\snn = $7.7-200 GeV. The numbers in parentheses represent the uncertainties.}
\begin{ruledtabular}
\begin{tabular}{cccccc}
\multirow{2}*{$\snn$ (GeV)}	&	\multicolumn{5}{c}{Collision Centralities}\\
\cline{2-6}
&	$0-10\%$	&	$10-20\%$	&	$20-40\%$	&	$40-60\%$	&	$60-80\%$\\
\colrule
7.7	&	313(3)	&	226(8)	&	134(10)	&	58(10)	&	22(5)\\
11.5	&	313(4)	&	226(8)	&	135(10)	&	58(9)	&	20(7)\\
14.5	&	314(4)	&	226(8)	&	133(10)	&	57(9)	&	19(5)\\
19.6	&	314(3)	&	225(9)	&	133(10)	&	58(9)	&	20(6)\\
27	&	319(4)	&	234(9)	&	140(11)		&	61(10)	&	20(7)\\
39	&	317(4)	&	230(9)	&	137(11)		&	59(10)	&	20(6)\\
62.4	&	320(4)	&	232(8)	&	139(10)	&	60(10)	&	20(6)\\
200	&	325(4)	&	237(9)	&	143(11)		&	62(10)	&	21(6)\\
\end{tabular}
\end{ruledtabular}
\end{table}

Track selection criteria for all analyses are presented in Table~\ref{tab:3}. In order to suppress the admixture of tracks from secondary vertices, a requirement of less than 1 cm is placed on the distance of closest approach (DCA) between each track and the event vertex. Furthermore, tracks must have at least 20 points ($n$FitPts) used in track fitting out of the maximum of 45 hits possible in the TPC. To prevent multiple counting of split tracks, at least $52\%$ of the total possible fit points ($n$FitPoss) are required. A condition is placed on the number of $dE/dx$ points ($ndE/dx$) used to derive $dE/dx$ values. The results presented here are within rapidity $|y| <$ 0.3 and have the same track cuts for all energies.

\begin{table}[h!]
\caption{\label{tab:3}Track selection criteria at all energies.}
\begin{ruledtabular}
\begin{tabular}{ccccc}
$n$FitPts		&	$n$FitPts/$n$FitPoss	&	$ndE/dx$	&	DCA	&	$|y|$\\
\colrule
$\geqslant$ 20	&	$\geqslant$ 0.52	&	$\geqslant$ 10		&	$\leqslant$ 1 cm	&	$<  $ 0.3\\
\end{tabular}
\end{ruledtabular}
\end{table}

\subsection{\label{sec:level2-3}Particle identification}
\begin{figure*}[!htp]
\centering
\centerline{\includegraphics[width=\textwidth]{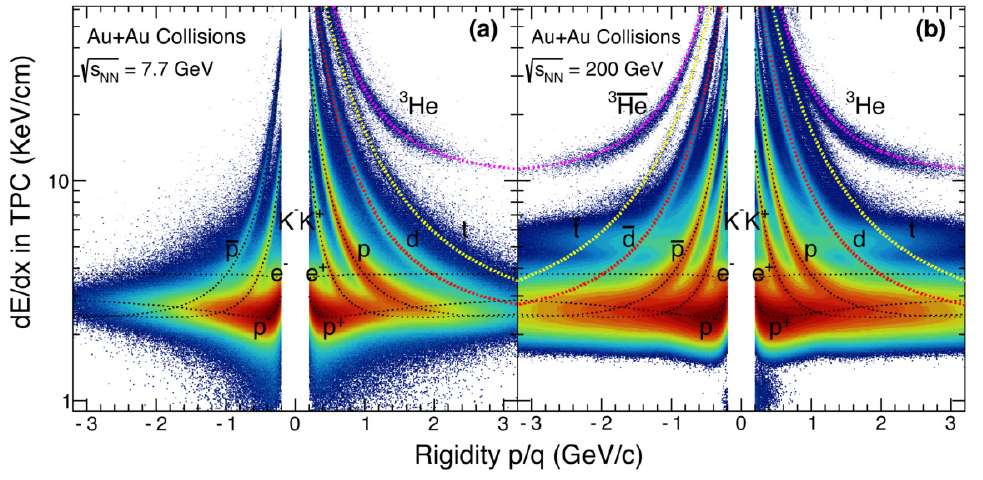}}
\caption{\label{fig:1}(Color online) Energy loss $dE/dx$ versus rigidity for TPC tracks from $0-80\%$ centrality Au+Au collisions at $\snn = $ 7.7 GeV (a) and 200 GeV (b) . The dashed lines represent a parametrization of the Bichsel function (see text for details) curve for different particles.}
\end{figure*}

\begin{figure}[!htp]
\centering
\centerline{\includegraphics[width=0.5\textwidth]{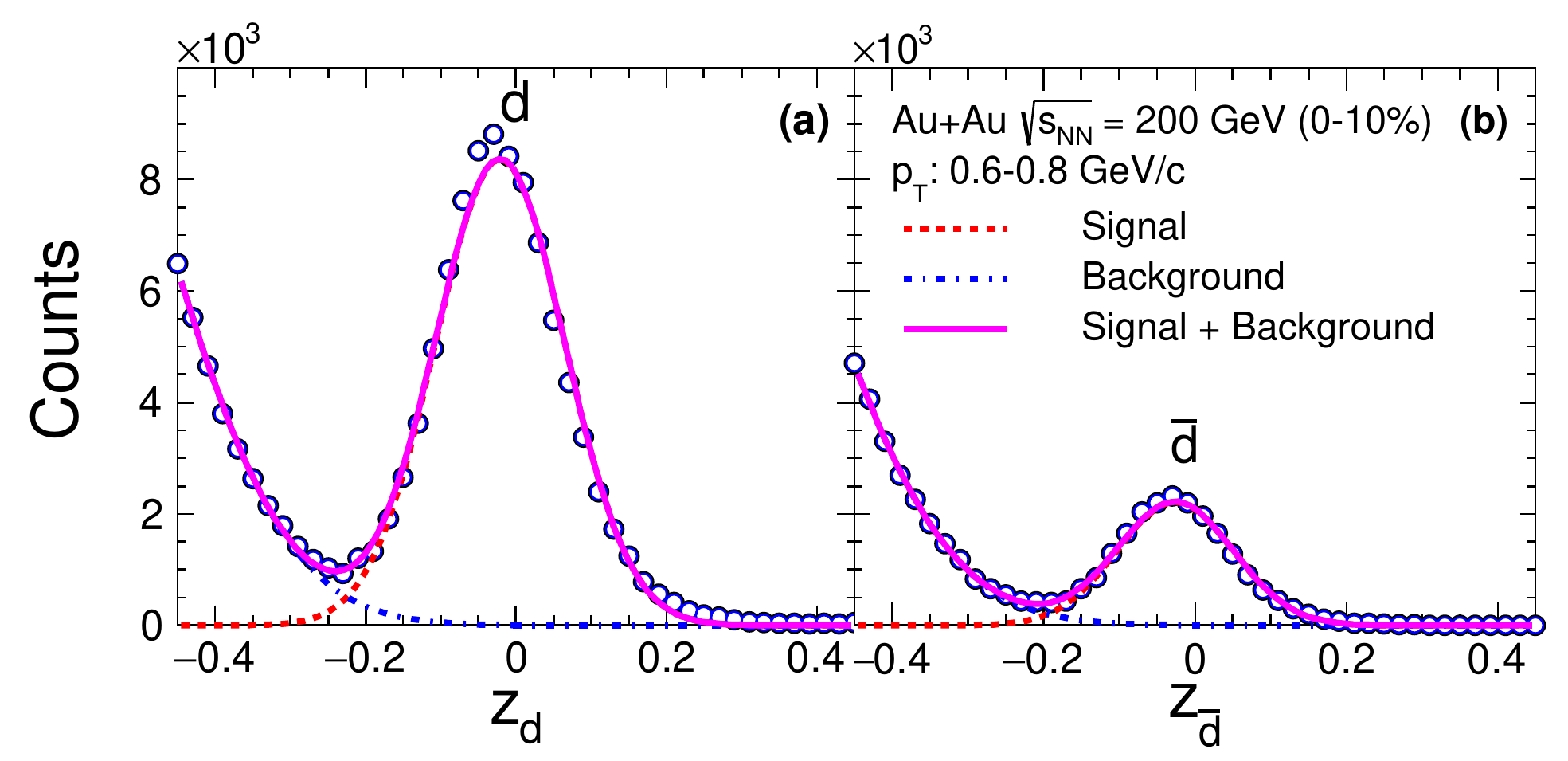}}
\caption{\label{fig:tpc} The $z(d)$ distributions for positively (a) and negatively (b) charged particles in the TPC at $p_{T} = 0.6-0.8$ \gc\ in $0-10\%$ centrality Au+Au collisions at $\snn = $ 200 GeV. The curves are two-Gaussian fits for signal and background. Uncertainties are statistical only and are smaller than the marker size.}
\end{figure}

Particle identification is mainly performed using the TPC. It is based on the measurement of the specific ionization energy deposit ($dE/dx$) of charged particles. Figure~\ref{fig:1} shows $dE/dx$ versus rigidity (momentum/charge, $p/q$) for TPC tracks from the $0-80\%$ centrality Au+Au collisions at $\snn = $ 7.7 and 200 GeV. The dotted curves represent a parametrization of the Bichsel function~\cite{BICHSEL2006154} for the different particle species. It is observed that the TPC can identify deuterons ($d$) and anti-deuterons ($\bar{d}$) at low momentum. The $dE/dx$ distribution for a fixed particle type is not Gaussian, hence a new variable $z$ is useful in order to have a proper deconvolution into Gaussian~\cite{Aguilar-Benitez1991}, which is defined as
\begin{equation}\label{zfun}
z=\text{ln}\left(\frac{\langle dE/dx\rangle}{\langle dE/dx\rangle_\text{B}}\right),
\end{equation}
where $\langle dE/dx\rangle_\text{B}$ is the Bichsel function for each particle species. The $z$ distributions for deuterons ($z_d$) and anti-deuteron ($z_{\bar{d}}$) are shown in Fig.~\ref{fig:tpc}, measured in $0-10\%$ centrality Au+Au collisions at $\snn$ = 200 GeV, for positively and negatively charged particles within the transverse momentum range $p_T = 0.6-0.8 $ \gc. To extract the raw yield of deuterons for $\pt < $ 1 \gc, a multi-Gaussian fit is applied to the $z$ distribution.

\begin{figure*}[!htp]
\centering
\centerline{\includegraphics[width=\textwidth]{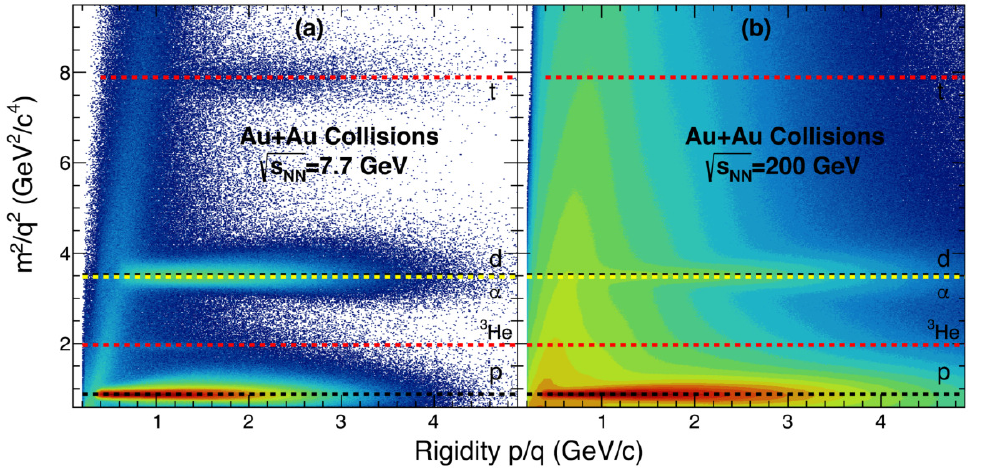}}
\caption{\label{fig:2} $m^2/q^2$ versus rigidity for ToF tracks from $0-80\%$ centrality Au+Au collisions at $\snn = $ 7.7 GeV (a) and 200 GeV (b). The dashed lines represent the $m^2/q^2$ values for different particles.}
\end{figure*}

\begin{figure}[!htp]
\centering
\centerline{\includegraphics[width=0.5\textwidth]{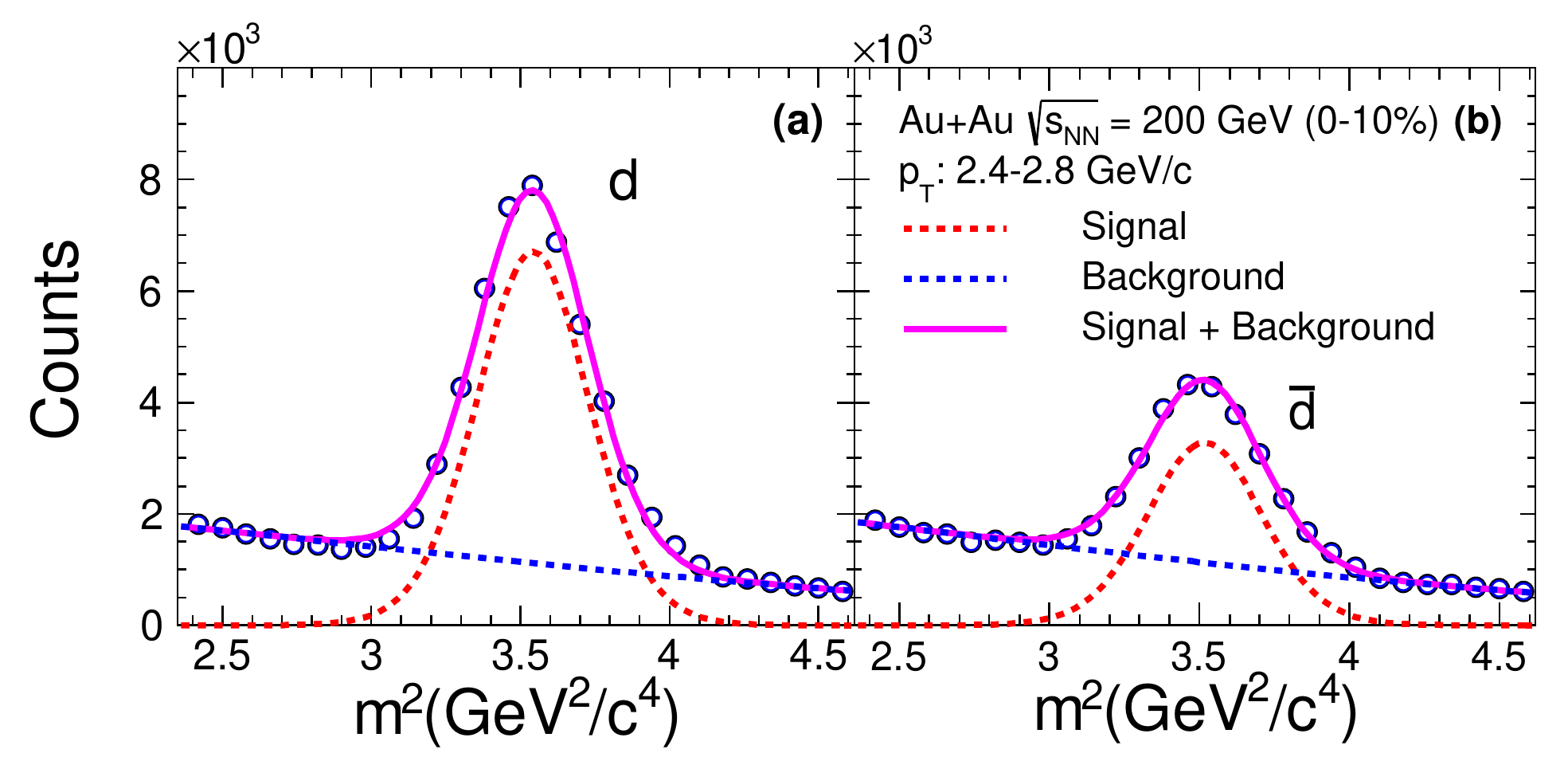}}
\caption{\label{fig:tof} The $m^2$ distributions for positively (a) and negatively (b) charged particles in the ToF at $p_{T} = 2.4-2.8$ \gc\ in the Au+Au $0-10\%$ centrality collisions at $\snn = $ 200 GeV. The curves are student's $t-$ and exponential fits for signal and background. Uncertainties are statistical only and are smaller than the point size.}
\end{figure}

The raw yield of deuterons above 1 \gc\ is obtained from the ToF data, using the mass square ($m^2$), defined as
\begin{equation}\label{zfun}
\frac{m^2}{q^2}=\frac{p^2}{q^2}\left(\frac{c^2t^2}{L^2}-1\right),
\end{equation}
where $t$, $L$, and $c$ are the flight time of a particle, track length, and speed of light. For the charged particles with more than unit charge, this definition is not the same as its real mass square. The mean value of $m^2$ ($q =1$) for deuteron is 3.52 GeV$^2/c^4$. Figure~\ref{fig:2} shows the $m^2/q^2$ versus rigidity for ToF tracks from the $0-80\%$ centrality Au+Au collisions at $\snn = $ 7.7 and 200 GeV. The dotted straight lines represent the $m^2/q^2$ for the different particle species. It can be observed that the ToF can extend the identification for deuterons and anti-deuterons up to $4\sim5$ GeV. The $m^2$ distributions of positively and negatively charged particles within the transverse momentum interval $p_T = 2.4-2.8 $ \gc\ measured in $0-10\%$ centrality Au+Au collisions at $\snn$ = 200 GeV, are shown in Fig.~\ref{fig:tof}. Since the $m^2$ distribution is not exactly Gaussian, the $m^2$ distribution is fitted with a student's $t$-function with an exponential tail for the signal~\cite{Pfanzagl:1996gf,PhysRevC.88.014902}.

\section{\label{sec:level3}Corrections and systematic uncertainties}
\subsection{\label{sec:level3.1}Corrections}
The final $\pt$ spectra of (anti-)deuterons are obtained by correcting the raw spectra for tracking efficiency and acceptance. These are determined by embedding tracks generated by Monte Carlo (MC) using GEANT3 of the STAR detector into real events at the raw data level~\cite{PhysRevLett.87.262302}. The ratio of the distribution of reconstructed and original MC tracks as a function of $\pt$ gives the efficiency $\times$ acceptance correction. Due to the unknown interactions of anti-nuclei with material, these processes are not included for anti-nuclei heavier than anti-protons in GEANT3. This lack of data on the anti-deuteron results in the calculated embedding efficiency being too high. This deficiency in the simulations is corrected for via an ``absorption correction" in STAR~\cite{PhysRevLett.87.262301}. A full detector simulation with GEANT4 was used, which has extensively validated cross-sections for light (anti)nuclei based on experimental data~\cite{Galoyan2013}.The loss of (anti)-nuclei due to interactions with the detector material within GEANT3 was then scaled to match the values from GEANT4. In this way a complete efficiency $\times$ acceptance correction in the relevant phase space is obtained.

ToF detector information is added to the information from TPC detector to give better particle identification at higher momenta. This requires an extra correction called the ToF matching efficiency. The ToF matching efficiency is defined as the ratio of the number of tracks matched in the ToF to the number of the total tracks in the TPC within the same acceptance, which is of the similar value to that of the protons~\cite{PhysRevC.96.044904}.

Low-momentum particles lose a considerable amount of energy while traversing the detector material. The track reconstruction algorithm takes into account the Coulomb scattering and energy loss, assuming the pion mass for each particle. Therefore, a track-by-track correction for the energy loss of heavier particles is needed. This correction is obtained from MC simulation using GEANT3, in which the $\pt$ difference of the reconstructed and the embedded particles is compared. The energy loss correction for deuterons is about $2\%$ at $\pt = $ 0.6 \gc\ and decreases with increasing $\pt$.

\begin{figure}[!htp]
\centering
\centerline{\includegraphics[width=0.5\textwidth]{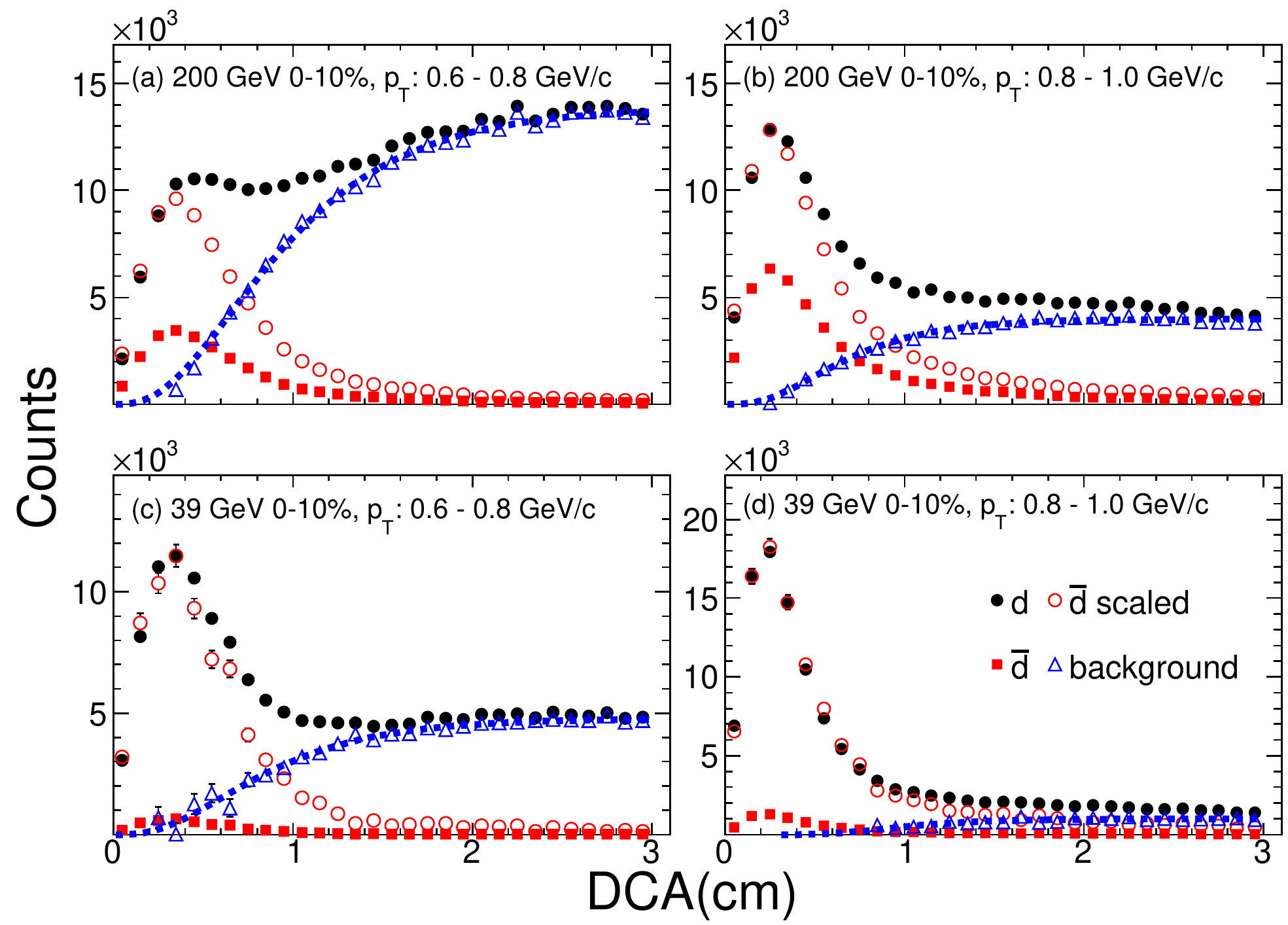}}
\caption{\label{fig:bp} DCA distributions of deuterons and anti-deuterons for $0.6<p_T<0.8$ GeV/$c$, and $0.8<p_T<1.0$ GeV/$c$ in Au+Au $0-10\%$ centrality collisions at $\snn = $ 200 GeV (a,b) and 39 GeV (c,d). Uncertainties are statistical only. The dashed curve is the fitted deuteron background. The red circles are the DCA distribution for anti-deuteron scaled up by the parameter $A$.}
\end{figure}

The so-called knock-out deuterons, due to interactions of energetic particles produced in collisions with detector material were observed in the STAR experiment. They are produced away from the primary interaction point and appear as a long tail in the DCA distribution of deuterons. The long and flat DCA tail in the deuteron distribution comes mainly from knock-out background deuterons. There are no knock-out anti-deuterons, hence the flat tail in the DCA distribution is absent. The knock-out deuterons can be determined by comparing the shapes of the DCA distribution of deuterons and anti-deuterons~\cite{PhysRevC.70.041901,PhysRevC.79.034909}. It is assumed that the shape of the background-subtracted deuteron DCA distribution is identical to that of the anti-deuteron. The DCA distribution of deuterons can be fitted by
\begin{equation}
N_d(\text{DCA}) = A\cdot N_{\bar{d}}(\text{DCA}) + N_d^{\text{back.}}(\text{DCA}),
\end{equation}
where $N_d^{\text{back.}}$ are knock-out deuterons with the form $N_d^{\text{back.}}=B\cdot\left(1-e^{\text{DCA}/C}\right)^D$. $A$, $B$, $C$, and $D$ are fit parameters. We used this functional form to fit the deuteron DCA distributions for every $\pt$ bin in each centrality at each energy to obtain the fraction of deuteron background. Figure~\ref{fig:bp} shows DCA distributions of deuterons and anti-deuterons for $0.6<p_T<0.8$ GeV/$c$, and $0.8<p_T<1.0$ GeV/$c$ in Au+Au $0-10\%$ centrality collisions at $\snn = $ 200 GeV and 39 GeV. Similarly to the results of protons~\cite{PhysRevC.96.044904}, the deuteron background fraction decreases with increasing $\pt$ and decreasing $\snn$. In $0-10\%$ centrality Au+Au collisions the background fraction at $\pt = 0.6-0.8\ $\gc\ is about $16\%$ for $\snn = $ 39 GeV and $33\%$ for 200 GeV with DCA $<$ 1 cm. These effects can be neglected when $\pt >$ 1.0 \gc.

\subsection{\label{sec:level3.2}Systematic uncertainties}

The point-to-point systematic uncertainties of the spectra are estimated by varying event and track selection and analysis cuts as well as by assessing sample purity from the $dE/dx$ measurement. In addition, fitting ranges of $z$(TPC) and $m^2$(ToF) are varied to estimate the systematic uncertainty on the extracted raw spectra. The estimated uncertainties are less than $4\%$ for all collision energies, in case of deuterons. These uncertainties increase with decreasing collision energy for anti-deuterons. For deuteron spectra, an additional systematic uncertainty appears due to background subtraction. This estimated uncertainty is $6-10\%$ at $\pt = $ 0.6-0.8 \gc\ and $2-5\%$ at $\pt = $ 0.8-1.0 \gc. A correlated overall systematic uncertainty of $5\%$ is estimated for all spectra and is dominated by uncertainties in the MC determination of the reconstruction efficiency. 

The $\pt$-integrated particle yields ($dN/dy$) and the average transverse momentum ($\langle \pt\rangle$) are calculated from the measured $\pt$ range and extrapolated to the unmeasured regions with individual Blast-Wave (BW) model fits~\cite{PhysRevC.48.2462}. The contribution to the yields from extrapolation is about $10-20\%$, which is from the low $p_T$ range. This model describes particle production properties by assuming that the particles are emitted thermally from an expanding source. The functional form of the model is given by
\begin{equation}\label{bw}
\frac{1}{2\pi p_T}\frac{d^2N}{dp_Tdy}\propto\int^R_0rdrm_TI_0\left(\frac{p_T\sinh\rho}{T}\right)K_1\left(\frac{m_T\cosh\rho}{T}\right),
\end{equation}
where $m_T=\sqrt{m^2+p_T^2}$ is the transverse mass ($m$ is the rest mass), $\rho = \tanh^{-1}\beta = \tanh^{-1}\left[\beta_s\left(r/R\right)^n\right]$ is the velocity profile, $I_0$ and $K_1$ are the modified Bessel functions, $r$ is the radial distance from the center of the thermal source in the transverse plane, $R$ ($=10fm$) is the radius of the thermal source, $\beta$ is the transverse expansion velocity, $\beta_s$ is the transverse expansion velocity at the surface, $n$ is the exponent of the velocity profile, and $T$ is the kinetic freeze-out temperature. 

The systematic uncertainties for $dN/dy$ and $\langle \pt\rangle$ are from cuts ($\sim 4\%$), tracking efficiency ($\sim 5\%$), energy loss ($\sim 2\%$). The extrapolation is an additional source of systematic uncertainty on $dN/dy$ and $\langle \pt\rangle$. This is estimated by comparing the extrapolation by different fit functions to the $\pt$ spectra. These functions are as follows:
\begin{eqnarray}
&&\textrm{Boltzmann} :\ \propto m_T \exp \left( -m_T/T \right)\nonumber\\
&&\textrm{double-exponential} :\ A_1\exp \left( -p^2_T/T_1^2 \right) + A_2\exp \left( -p^2_T/T_2^2 \right)\nonumber
\end{eqnarray}

This uncertainty is $5-10\%$ for deuterons at all energies and anti-deuterons at $\snn \geqslant $27 GeV, and of the order of $15\%$, $30\%$, and $50\%$ for anti-deuterons at $\snn = $ 19.6, 14.5, and 11.5 GeV, respectively. The total systematic uncertainties are calculated as quadrature sums of all the components discussed above.

\section{\label{sec:level4}Results and discussions}
\subsection{\label{sec:level4-1}Transverse momentum spectra}
\begin{figure*}[!htp]
\centering
\includegraphics[width=0.5\textwidth]{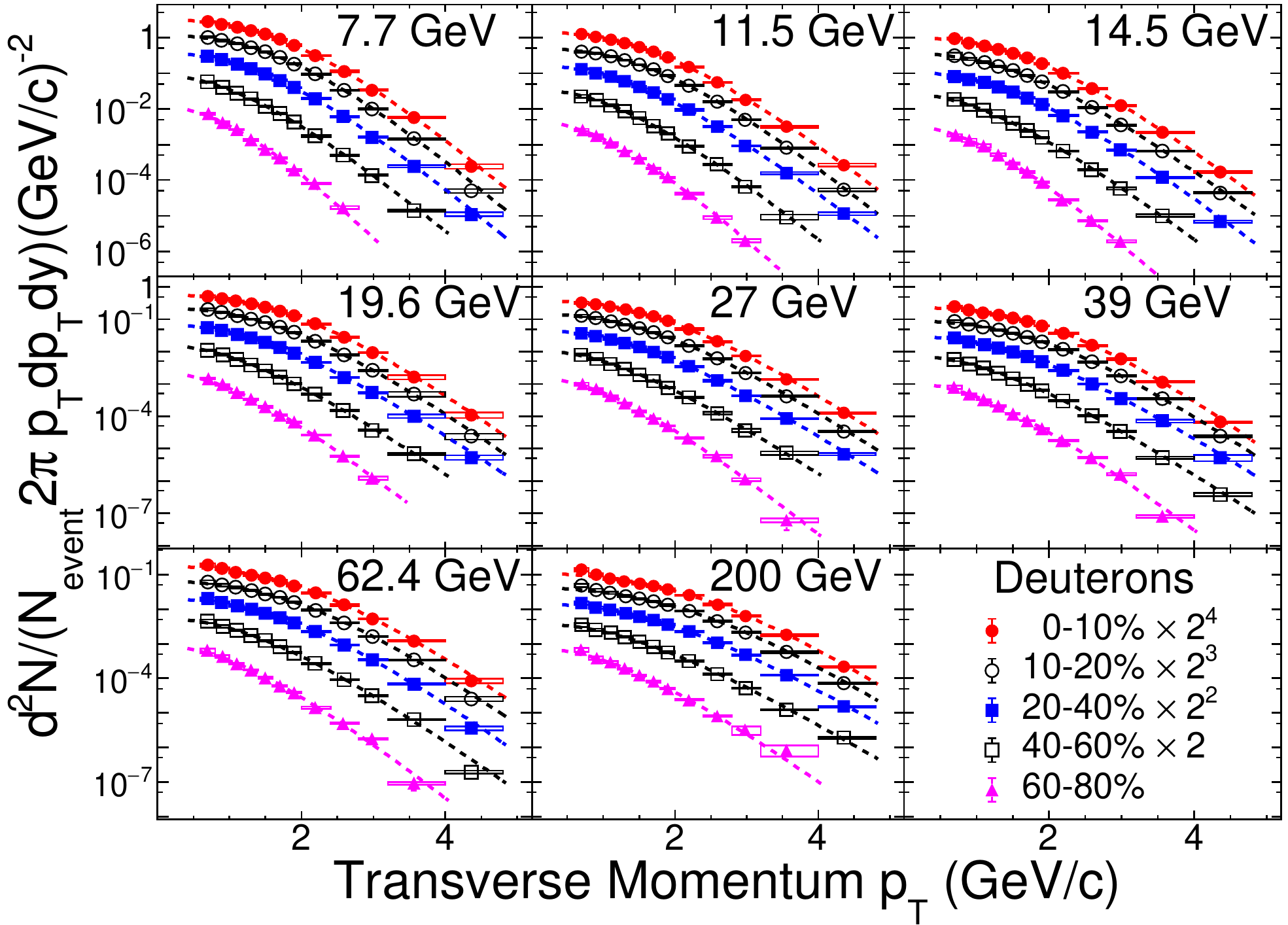}
\includegraphics[width=0.5\textwidth]{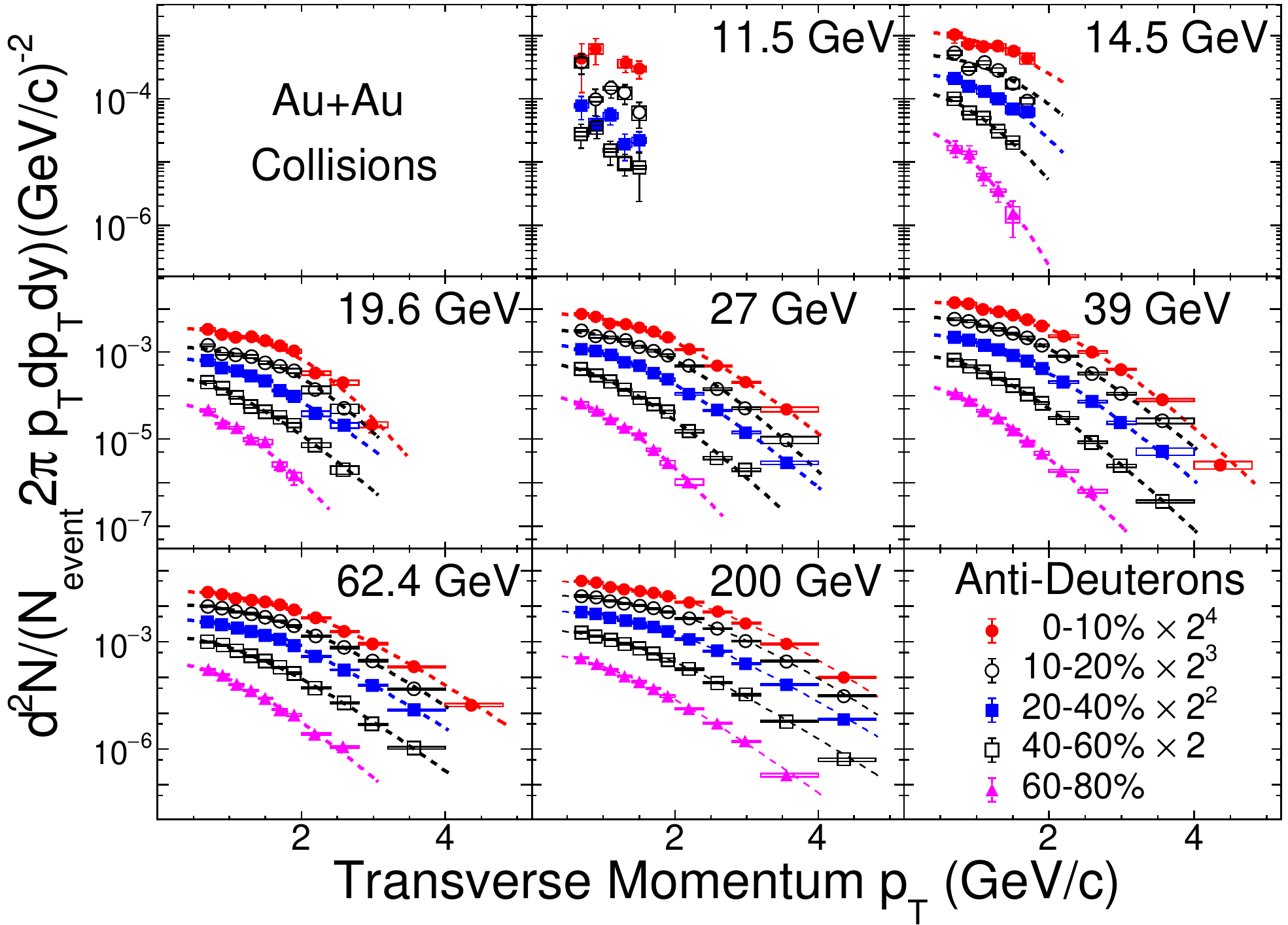}
\caption{\label{fig:tmd} Mid-rapidity $(|y| < 0.3)$ transverse momentum spectra for deuterons and anti-deuterons in Au+Au collisions at $\snn =$ 7.7, 11.5, 14.5, 19.6, 27, 39, 62.4, and 200 GeV for $0-10\%$, $10-20\%$, $20-40\%$, $40-60\%$, and $60-80\%$ centralities. There are not enough candidates to obtain the anti-deuterons $p_T$ spectra at $\snn =$ 7.7 GeV. The dashed-lines are the results of the Blast-Wave fits to each distribution with the profile parameter $n$=1, see discussions in the text. The statistical and systematical uncertainties are shown as vertical-lines and boxes, respectively. Horizontal uncertainties reflect the $p_T$ bin width used in the analysis. The spectra for different centralities are scaled for clarity.}
\end{figure*}

Figure~\ref{fig:tmd} shows mid-rapidity $(|y| < 0.3)$ transverse momentum spectra for deuterons (left panel) and anti-deuterons (right panel) in Au+Au collisions at $\snn = $ 7.7, 11.5, 14.5, 19.6, 27, 39, 62.4, and 200 GeV for $0-10\%$, $10-20\%$, $20-40\%$, $40-60\%$, and $60-80\%$ centralities. At 7.7 GeV, the statistics and the yield of the anti-deuterons are too small to obtain the $p_T$ spectra. The dashed lines are the result of the Blast-Wave fits to each distribution with the profile parameter $n$=1 (Eq.~(\ref{bw})). The $\pt$ spectra show a clear evolution, becoming somewhat softer from central to peripheral collisions. A similar behavior is observed for kaons and protons~\cite{PhysRevC.79.034909}.

\subsection{\label{sec:level4-2}Average transverse momenta $\langle p_T\rangle$}
\begin{figure}[!htp]
\centering
\centerline{\includegraphics[width=0.5\textwidth]{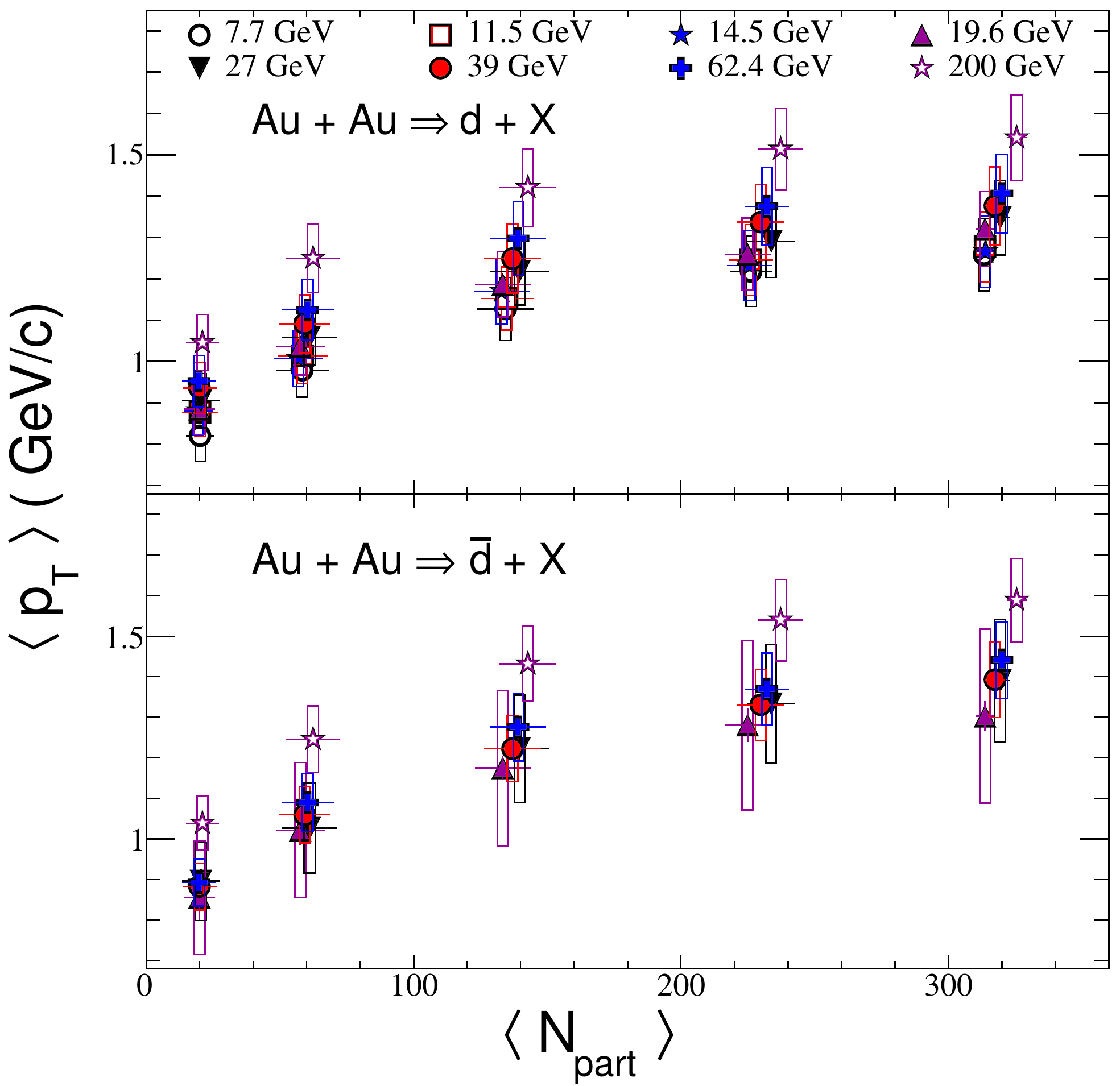}}
\caption{\label{fig:mpt} Centrality dependence of $\langle \pt \rangle$ of deuterons (top panel) and anti-deuterons (bottom panel) in Au+Au collisions at $\snn = $ 7.7, 11.5, 14.5, 19.6, 27, 39, 62.4, and 200 GeV. The $\bar{d}$ results at 11.5 and 14.5 GeV are not presented due to their large uncertainty. The statistical and systematical uncertainties are shown as vertical lines and boxes, respectively.}
\end{figure}

Figure~\ref{fig:mpt} shows $\langle \pt \rangle$ as a function of $\langle N_{\text{part}}\rangle$ for deuterons and anti-deuterons in Au+Au collisions at $\snn = $ 7.7, 11.5, 14.5, 19.6, 27, 39, 62.4, and 200 GeV. The dependence of $\langle \pt \rangle$ on $\langle N_{\text{part}}\rangle$ for deuterons and anti-deuterons are similar to those for $\pi^{\pm}$, $K^{\pm}$, protons, and anti-protons~\cite{PhysRevC.79.034909}. An increase in $\langle \pt \rangle$ with increasing number of participants is observed at all collision energies. A slight increase with increasing collision energies is also found, which suggests that the average radial flow increases with collision energy and centrality.

\subsection{\label{sec:level4-3}Particle yields $dN/dy$}
\begin{figure}[!htp]
\centering
\centerline{\includegraphics[width=0.5\textwidth]{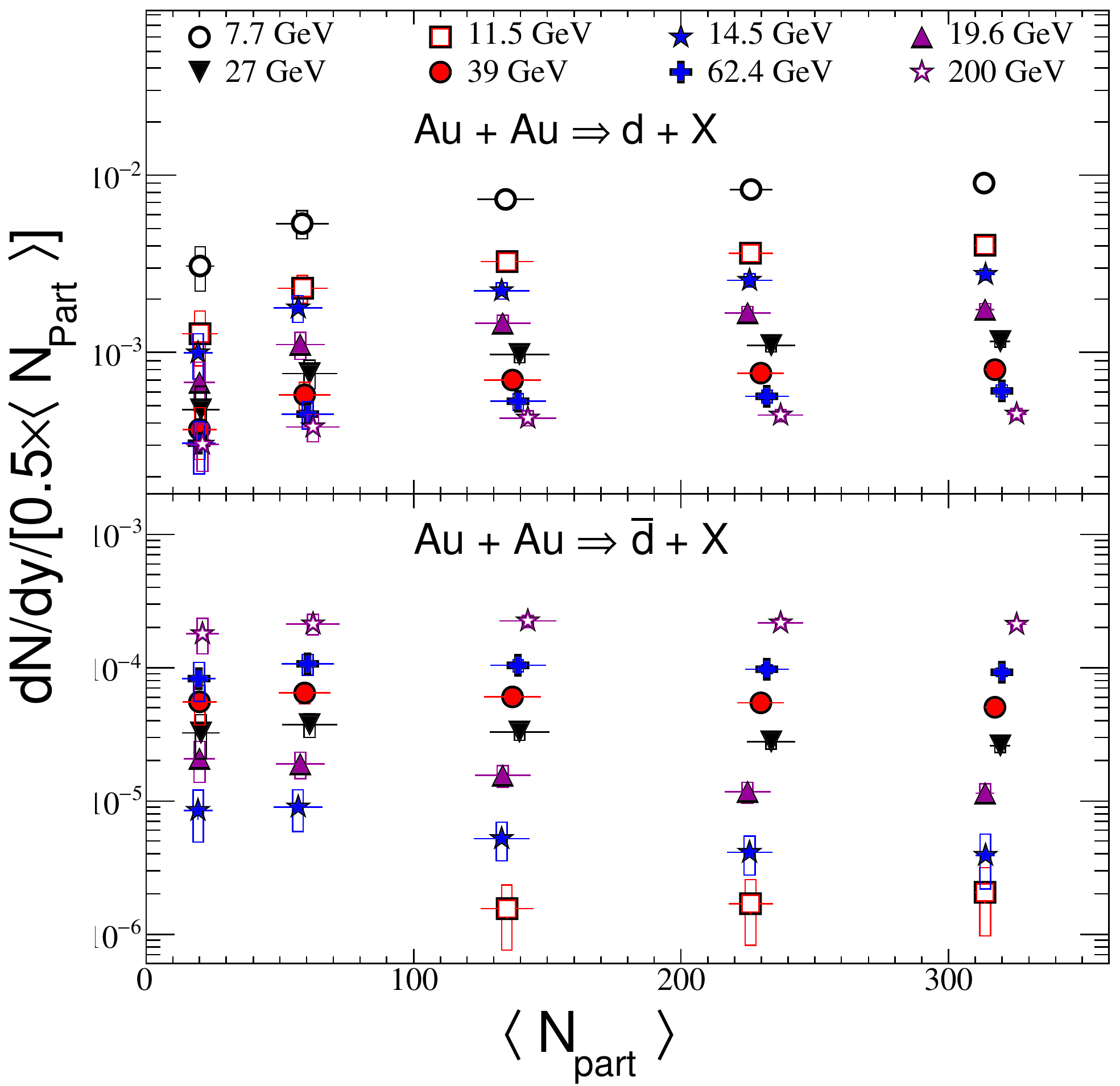}}
\caption{\label{fig:dndy} Centrality dependence of $dN/dy$ normalized by $0.5\langle N_{\text{part}}\rangle$ of deuterons (top panel) and anti-deuterons (bottom panel) in Au+Au collisions at $\snn = $ 7.7, 11.5, 14.5, 19.6, 27, 39, 62.4, and 200 GeV. The statistical and systematical uncertainties are shown as vertical lines and boxes, respectively.}
\end{figure}

Figure~\ref{fig:dndy} shows the centrality dependence of deuterons and anti-deuterons rapidity density ($dN/dy$) at mid-rapidity $(|y| < 0.3)$, normalized by $0.5\langle N_{\text{part}}\rangle$ for $\snn = $ 7.7, 11.5, 14.5, 19.6, 27, 39, 62.4, and 200 GeV. The production of deuterons is expected to be mainly from two processes, pair production and baryon stopping. As collision energy increases, contributions from pair production increase while the contributions from baryon stopping decrease. The deuteron yield decreases from 7.7 GeV to 200 GeV implying that in this energy window stopping plays a more important role than pair-production for deuterons. On the other hand, due to pair production being the only source for anti-nucleon and anti-deuteron production, the yields of anti-deuterons increase as the energy increases. Furthermore, the normalized yields for deuterons decrease from central to peripheral collisions suggesting the effect of baryon stopping is stronger in more central collisions. However, the centrality dependence of normalized yields for anti-deuteron is weaker.

\subsection{\label{sec:level4-4}Particle ratios}
\begin{figure}[!htp]
\centering
\centerline{\includegraphics[width=0.5\textwidth]{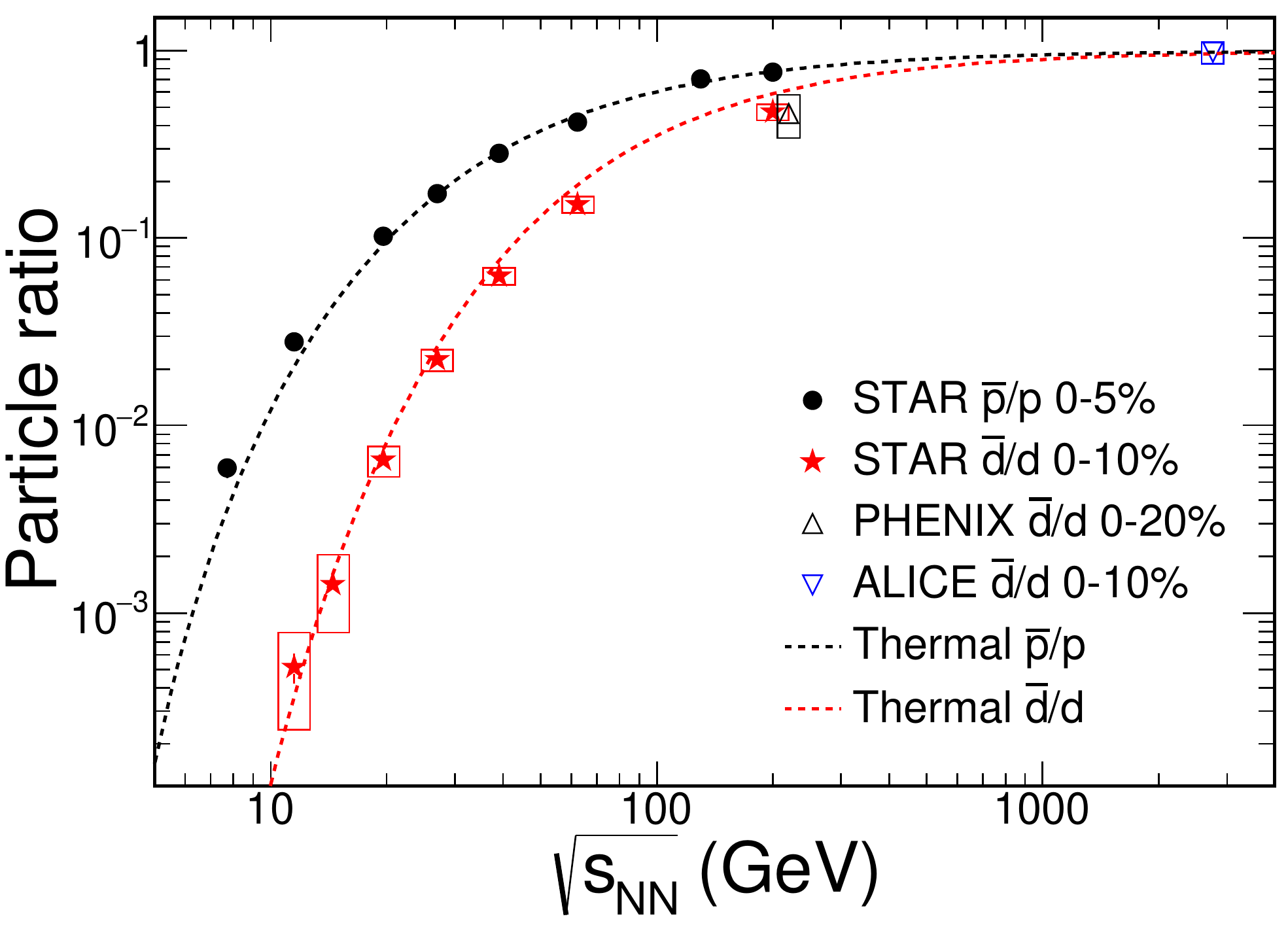}}
\caption{\label{fig:dbd} Energy dependence of $\bar{p}/p$, $\bar{d}/d$ ratios from Au+Au collisions at RHIC~\cite{PhysRevC.96.044904}. The PHENIX and ALICE data points are shown as triangles and inversed triangles~\cite{PhysRevLett.94.122302,ALICE}. The curves are thermal model results as described in the text.}
\end{figure}

Figure~\ref{fig:dbd} shows the anti-particle over particle ratios ($\bar{p}/p$ from $0-5\%$ centrality, $\bar{d}/d$ from $0-10\%$ centrality) in Au+Au collisions at $\snn = $ 7.7, 11.5, 14.5, 19.6, 27, 39, 62.4, and 200 GeV. For comparison, the PHENIX and ALICE data points are shown. Both ratios $\bar{p}/p$ and $\bar{d}/d$ approach unity at higher collision energies. This can be attributed to the decreasing net baryon density at mid-rapidity, as well as due to $p(\bar{p})$, $d(\bar{d})$ production becoming dominated by pair production.

In the framework of statistical thermal models~\cite{PhysRevC.73.034905} the particle multiplicity from a source of volume $V$ and chemical freeze-out temperature $T$ is given by

\begin{equation}\label{thermaleq}
N_i=\frac{g_iV}{\pi^2}m_i^2TK_2(m/T)\exp(\mu_i/T),
\end{equation}
where $g_i$, $m_i$, and $\mu_i$ are the degeneracy, particle mass, and chemical potential of particle species $i$ respectively. This formula is valid in the Boltzmann approximation, which is reasonable for all hadrons and light nuclei. The chemical potential can be expressed as $\mu_i=B_i\mu_B+S_{i}\mu_S+Q_i\mu_Q$, where $B_i$, $S_i$, and $Q_i$ are the baryon number, strangeness and charge, respectively, of particle species $i$, and $\mu_B$, $\mu_S$, and $\mu_Q$ are the corresponding chemical potentials for these conserved quantum numbers.

Results calculated by a statistical thermal model using the parametrizations for $T$ and $\mu_B$ established in~\cite{1742-6596-779-1-012012} are shown in Fig.~\ref{fig:dbd}. The thermal model can describe the $\bar{p}/p$ and $\bar{d}/d$ ratios over a very wide energy range. The $\bar{p}/p$ ratio is calculated for measured inclusive protons. The difference of weak-decay fractions for $p$ and $\bar{p}$ reaches a maximum around $\snn = $ 6 GeV~\cite{Yu2019}, which might be the reason of the deviation between measured and model $\bar{p}/p$ ratios at low energies.

\begin{figure}[!htp]
\centering{}
\centerline{\includegraphics[width=0.5\textwidth]{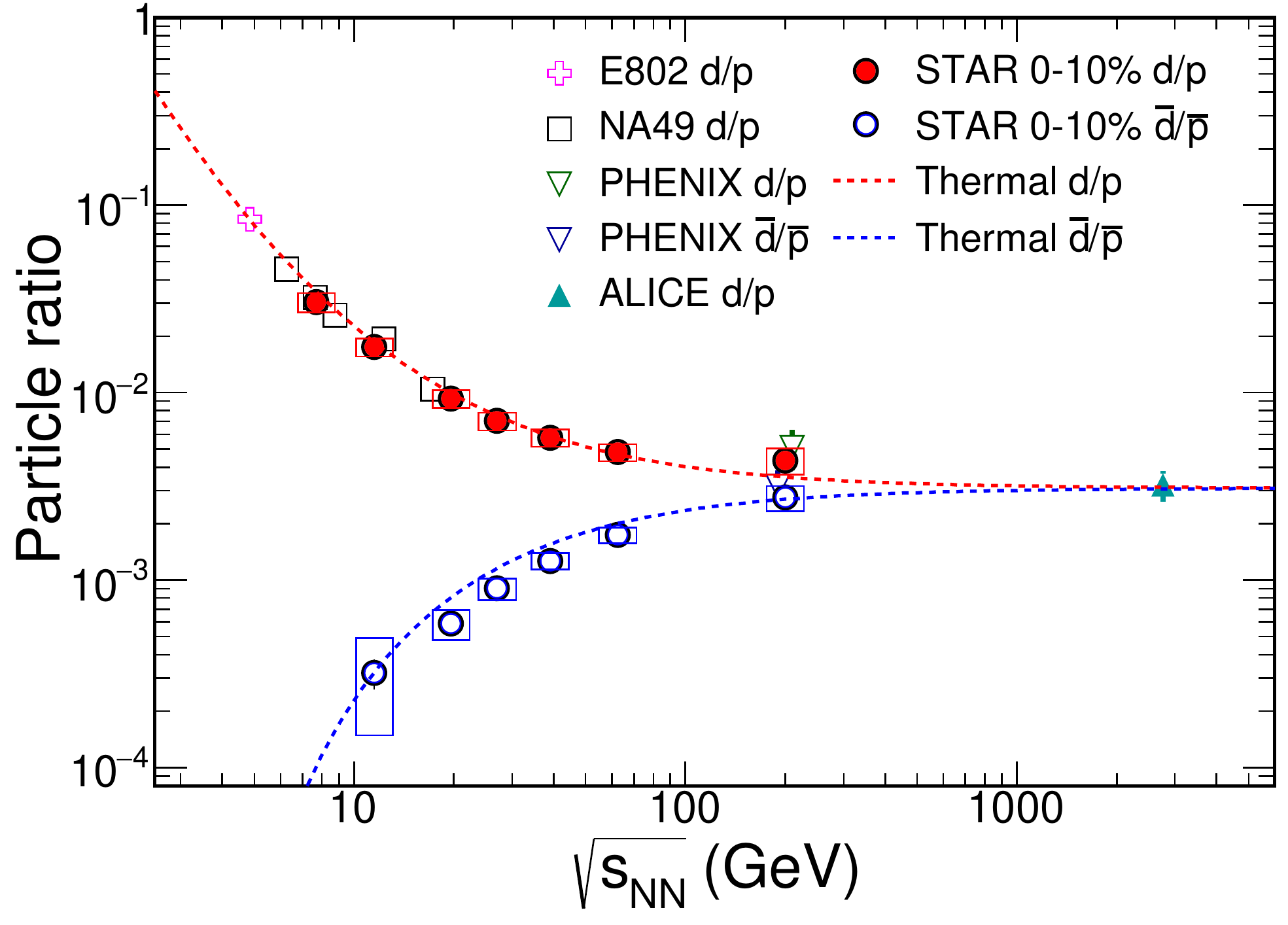}}
\caption{\label{fig:dpratio} Energy dependence of $d/p$ and $\bar{d}/\bar{p}$ yield ratios. The curves represent the thermal model results as described in the text. The symbols represent measured data~\cite{E802,NA49,PhysRevLett.94.122302,ALICE}.}
\end{figure}

Figure~\ref{fig:dpratio} shows the energy dependence of $d/p$ and $\bar{d}/\bar{p}$ yield ratios for the most central collision and are compared with those from E802~\cite{E802}, NA49~\cite{NA49}, PHENIX~\cite{PhysRevLett.94.122302}, and ALICE~\cite{ALICE}. The $p(\bar{p})$ yield is corrected by weak-decay feed-down from strange baryons~\cite{PhysRevLett.121.032301}. The $d/p$ ratios decrease and $\bar{d}/\bar{p}$ increase with increasing $\snn$ and both converge to the same value of about $3.6\times10^{-3}$ at LHC energy where the chemical potential is consistent with zero and hence the ratios are only determined by the chemical freeze-out temperature. Predictions by the statistical thermal model for $d/p$ and $\bar{d}/\bar{p}$ yield ratios are also shown in Fig.~\ref{fig:dpratio} by dashed curves and are in agreement with experimental data.

\begin{figure}[!htp]
\centering
\centerline{\includegraphics[width=0.5\textwidth]{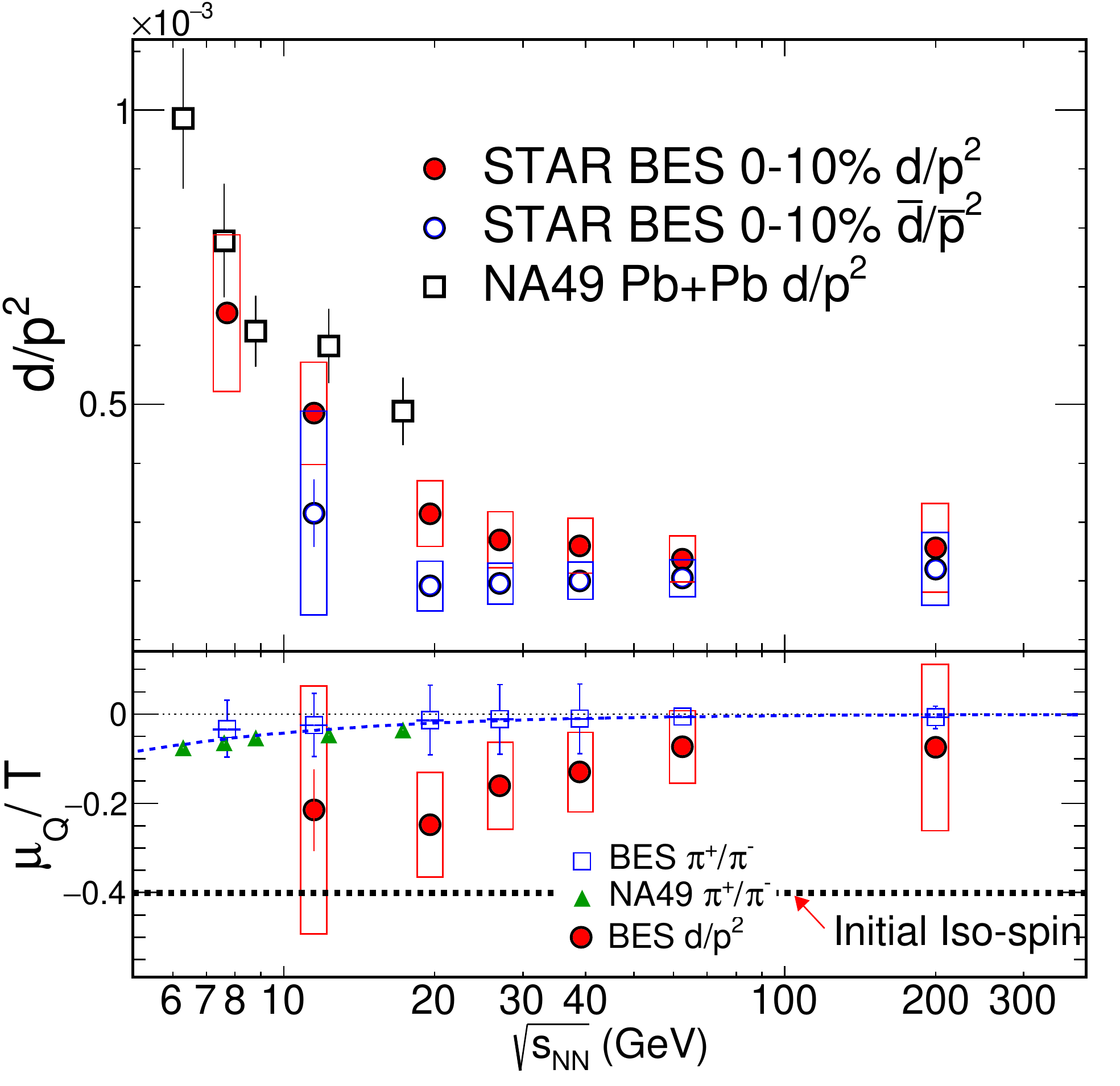}}
\caption{\label{fig:mui} Top panel : energy dependence of $d/p^{2}$, $\bar{d}/\bar{p}^2$ yield ratios at mid-rapidity for Au+Au top $10\%$ centrality collisions. Bottom panel : energy dependence of $\mu_Q/T$ from $d/p^{2}$($\bar{d}/\bar{p}^2$) and $\pi^+/\pi^-$ yield ratios. The blue dashed line is fit to the $\mu_Q/T$ from $\pi^+/\pi^-$ yield ratios by the function described in the text.}
\end{figure}

Figure~\ref{fig:mui} shows the energy dependence of the $\mu_Q/T$ values, which are related to the isospin effects in the collision system. These can be obtained by the $\pi^{+}/\pi^{-}$ ratio and $(\bar{d}/\bar{p}^2)/(d/p^2)$ from Eq.~(\ref{thermaleq}),

\begin{eqnarray}\label{piratio}
\frac{\mu_Q}{T}&=&\frac{1}{2}\ln\left(\frac{\pi^+}{\pi^-}\right),\\
\frac{\mu_Q}{T}&=&\frac{1}{2}\ln\left(\frac{\bar{d}/\bar{p}^2}{d/p^2}\right).
\end{eqnarray}
The deviation from Bose-Einstein distribution for pions in Eq.~(\ref{piratio}) can be neglected if $T < 180$ MeV and $\mu_Q/T > - 0.4$~\cite{Yu2019}. The energy dependence of $\bar{d}/\bar{p}^2$, $d/p^2$ yield ratios for top $10\%$ centrality Au+Au collisions are presented in the top panel of Fig.~\ref{fig:mui}. The NA49 results~\cite{NA49} are shown in this figure, they are consistent with STAR BES data, within uncertainties. The $\mu_Q/T$ extracted from BES $(\bar{d}/\bar{p}^2)/(d/p^2)$, $\pi^{+}/\pi^{-}$ data, and NA49 $\pi^{+}/\pi^{-}$ ratios are shown in the bottom panel of Fig.~\ref{fig:mui}. The $\mu_Q/T$ increases with $\snn$ and reaches zero at high collision energy, which suggests that the isospin effect is smaller at higher collision energies and most of the particles are from pair production. The $\mu_Q/T$ values extracted from $\pi^{+}/\pi^{-}$ are systematically larger than those from $(\bar{d}/\bar{p}^2)/(d/p^2)$ at small $\snn$ although the uncertainties in the values extracted from $(\bar{d}/\bar{p}^2)/(d/p^2)$ are large. The reasons for these lower $\mu_Q/T$ values could be due to the contribution of deuterons from nuclear fragments at low collision energy or non-thermal effects.

\subsection{\label{sec:level4-5}Coalescence}
\begin{figure}[!htp]
\centering
\centerline{\includegraphics[width=0.5\textwidth]{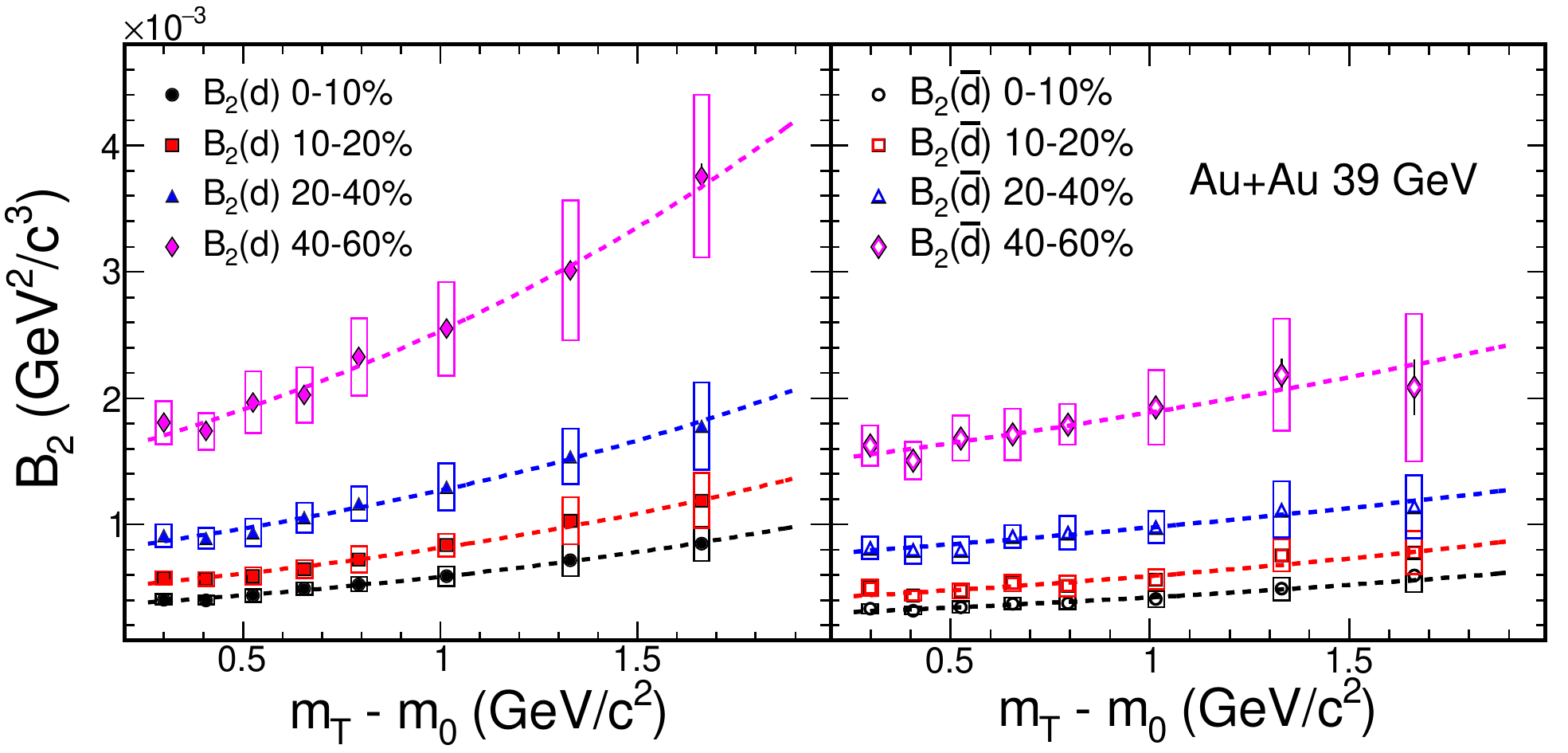}}
\caption{\label{fig:b2pt} Coalescence parameter $B_2$ as a function of $m_T-m_0$ for deuterons (left panel) and anti-deuterons (right panel) from $0-10\%$, $10-20\%$, $20-40\%$, and $40-60\%$ centrality at $\snn =$ 39 GeV Au+Au collisions. The boxes show systematic uncertainty and vertical lines show the statistical uncertainty separately. The dashed lines represent exponential fits.}
\end{figure}

The coalescence parameter $B_2$ (Eq.~(\ref{b2eq})) is studied using a combined analysis with the deuteron and proton spectra~\cite{PhysRevLett.121.032301,PhysRevC.69.034909}, which are corrected for weak decay feed-down. Figure~\ref{fig:b2pt} shows $B_2$ as a function of $m_T-m_0$ for deuterons and anti-deuterons from $\snn = 39$ GeV Au+Au collisions. It is found that the coalescence parameters $B_2$ increase with increasing transverse mass, which is a reasonable expectation based on the correlation volume increasing with decreasing $m_T$, leading to a higher coalescence probability for larger values of $m_T$. A decrease of $B_2$ with increasing centrality is found, which may be attributed to the volume of the collision system being larger in central collisions. The shape of $B_2$ can be described by an exponential form~\cite{Bearden2002,PhysRevC.83.044906} (as shown in Fig.~\ref{fig:b2pt}).
\begin{equation}
B_2=a\cdot\exp{\left[b(m_T-m)\right]},
\end{equation}
where $a$ denotes the coalescence parameter at $\pt = 0$, and $b$ is connected to the difference between the slope parameters of the spectra for deuterons and protons. The parameters for $\snn = 39$ GeV are listed in Table~\ref{tab:4}. It is found the $B_2$ values for deuterons are systematically larger than those for anti-deuterons in certain collision centrality at $\snn = 39$ GeV, which might signal that the effective distances of protons and neutrons are smaller than those of anti-protons and neutrons at 39 GeV as $B_2$ is inversely proportional to the correlation volume as shown by Eq.~(\ref{b2vsv}).

\begin{table}[h!]
\caption{\label{tab:4}The parameters $a$ and $b$ from exponential fitting of coalescence $B_2$ at $\snn =$ 39 GeV.}
\begin{ruledtabular}
\begin{tabular}{ccccc}
Centrality &	\multicolumn{2}{c}{$B_2(d)$}	&	\multicolumn{2}{c}{$B_2(\bar{d})$}\\	
\cline{2-5}
($\%$)	&	$a\times10^4$		&	$b$&	$a\times10^4$&	$b$\\
		&	(GeV$^2/c^3$)		&	($c^2$/GeV)	&	(GeV$^2/c^3$)		&($c^2$/GeV)	\\
\colrule
0-10		&	3.3$\pm$0.3	&	0.56$\pm$0.11	&	2.8$\pm$0.3	&	0.42$\pm$0.14\\
10-20	&	4.6$\pm$0.4	&	0.58$\pm$0.11	&	3.9$\pm$0.4	&	0.42$\pm$0.13\\
20-40	&	7.4$\pm$0.6	&	0.54$\pm$0.11	&	7.3$\pm$0.7	&	0.30$\pm$0.13\\
40-60	&	14.4$\pm$1.2	&	0.56$\pm$0.11	&	14.3$\pm$1.4	&	0.28$\pm$0.14\\
\end{tabular}
\end{ruledtabular}
\end{table}

\begin{figure}[!htp]
\centering
\centerline{\includegraphics[width=0.5\textwidth]{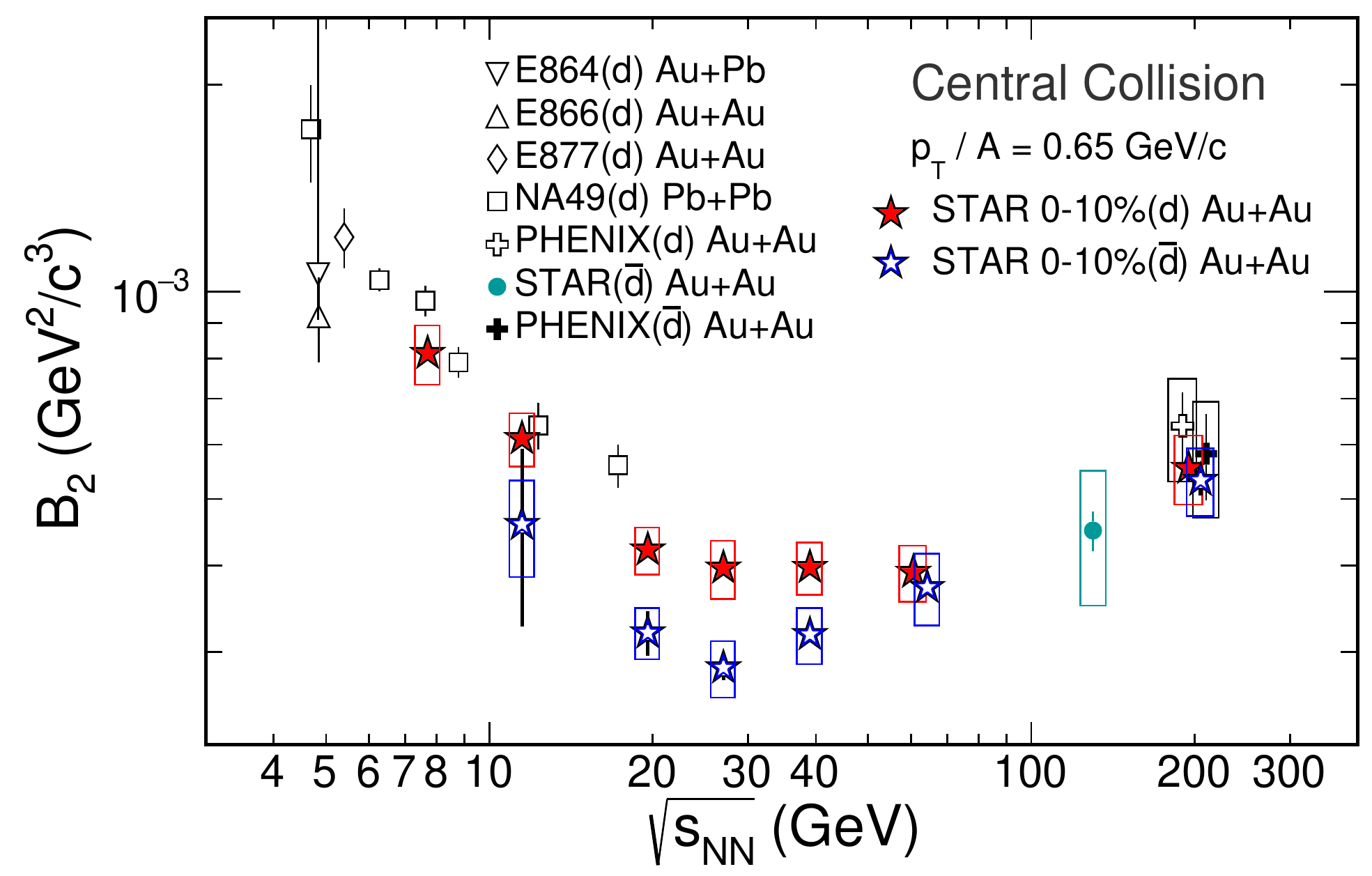}}
\caption{\label{fig:b2} Energy dependence of the coalescence parameter for $B_2(d)$ and $B_2(\bar{d})$ at $p_T/A$ = 0.65 GeV/$c$ from Au+Au collisions at RHIC. For comparison, results from 
AGS~\cite{E878,E864-1,E864-2}, SPS~\cite{Bearden2002,NA52,NA49} ($0-7\%$ and $0-12\%$ collision centralities), RHIC~\cite{PhysRevLett.87.262301,PhysRevLett.94.122302} ($0-18\%$ and $0-20\%$ collision centrality for $\snn =$ 130 GeV and 200 GeV) are also shown.}
\end{figure}

In Fig.~\ref{fig:b2}, we compare the collision energy dependence of $B_2$ at $\pt/A=0.65$ \gc\ in 0-10\% centrality Au+Au collisions at RHIC, as well as data from 
AGS~\cite{E878,E864-1,E864-2}, SPS~\cite{Bearden2002,NA52,NA49} ($0-7\%$ and $0-12\%$ collision centralities), RHIC~\cite{PhysRevLett.87.262301,PhysRevLett.94.122302} ($0-18\%$ and $0-20\%$ collision centrality for $\snn =$ 130 GeV and 200 GeV). The $B_2$ value at $\snn = $2.76 TeV from ALICE $0-20\%$ collision centrality of approximately $4 \times 10 ^{-4} \text{GeV}^2/c^3$~\cite{ALICE} is slightly lower than the measurement at RHIC. At energies below $\snn = 20$ GeV, the coalescence parameters $B_2$ decrease as a function of increasing collision energy, which implies that the overall size of the emitting source of nucleons increases with the collision energy. When $\snn >$ 20 GeV, the rate of decrease seems to change and saturate up to 62.4 GeV, which might imply a dramatic change of the equation of state of the medium in those collisions. The $B_2$ from 200 GeV is found to be larger than the BES saturation values, which needs further studies. The $B_2$ values for anti-deuterons are systematically lower than those for deuterons, which implies that the overall size of the emitting source of anti-baryons is larger than that of baryons. Again, at 200 GeV, the $B_2$ values for deuterons and anti-deuterons are the same within uncertainties. The similarity reflects the characteristics of pair-production. At lower collision energies, more and more stopped nucleons move into the mid-rapidity region, which suppresses the probability for the production for anti-deuterons. As a result, the $B_2$ values for anti-deuterons are reduced. The separation of $B_2$ between deuterons and anti-deuterons should increase as collision energy decreases. This will be tested in the future high statistics RHIC BES-II program, where the spectra at $\snn = $ 7.7, 9.2 and 11.5 GeV will be obtained with high precision.

\section{\label{sec:level5}Conclusion}
In conclusion, we have presented systematic studies of deuteron and anti-deuteron production in Au+Au collisions at $\snn = $ 7.7, 11.5, 14.5, 19.6, 27, 39, 62.4, and 200 GeV. The mid-rapidity yields $dN/dy$ show the effects of baryon stopping at lower collision energies. At higher collision energies, the pair production mechanism dominates the particle production. The anti-baryon to baryon yield ratios, and the $d/p$ yield ratio can be well reproduced by the thermal model. The $\mu_Q/T$ values extracted from $d/p^2$ ratios are systematically smaller than those from $\pi^+/\pi^-$, which may suggest that some of the observed deuterons are from the nuclear fragmentation. Two interesting new features are observed for the coalescence parameter $B_2$: ($i$) The values of $B_2$ for deuterons decrease as collision energy increases and seem to reach a minimum at about $\snn = 20 - 40$ GeV, indicating a change in the equation of state; ($ii$) $B_2$ values for anti-deuterons are found to be less than those for deuterons at collision energies below 62.4 GeV implying that the overall size of the emitting source of anti-baryons is larger than that of baryons at low collision energy.

\begin{acknowledgments}
We thank the RHIC Operations Group and RCF at BNL, the NERSC Center at LBNL, and the Open Science Grid consortium for providing resources and support. This work was supported in part by the Office of Nuclear Physics within the U.S. DOE Office of Science, the U.S. National Science Foundation, the Ministry of Education and Science of the Russian Federation, National Natural Science Foundation of China, Chinese Academy of Science, the Ministry of Education and Science of the Russian Federation, National Natural Science Foundation of China, Chinese Academy of Science, the Ministry of Science and Technology of China (973 Programme No. 2015CB856900) and the Chinese Ministry of Education, the National Research Foundation of Korea, GA and MSMT of the Czech Republic, Hungarian National Research, Development and Innovation Office (FK-123824), New National Excellency Programme of the Hungarian Ministry of Human Capacities (UNKP-18-4), Department of Atomic Energy and Department of Science and Technology of the Government of India; the National Science Centre of Poland, National Research Foundation, the Ministry of Science, Education and Sports of the Republic of Croatia, RosAtom of Russia and German Bundesministerium fur Bildung, Wissenschaft, Forschung and Technologie (BMBF) and the Helmholtz Association. 
\end{acknowledgments}
\bibliography{deuteron}
\end{document}